\documentclass[num-refs]{nbdt-article}

\definecolor{black}{rgb}{0.0, 0.0, 0.0}
\newcommand{\edit}[1]{\textcolor{black}{#1}}

\papertype{Original Article}
\paperfield{Commentary}

\title{Strong and weak principles of neural dimension reduction}

\author[1]{Mark D. Humphries}

\affil[1]{School of Psychology, University of Nottingham, UK}

\corraddress{School of Psychology, University of Nottingham, UK}
\corremail{mark.humphries@nottingham.ac.uk}

\fundinginfo{The time needed to think these thoughts was made available thanks to the generous support of the Medical Research Council grants MR/J008648/1, MR/P005659/1, and MR/S025944/1.}

\runningauthor{Mark D Humphries}

\begin{document}
\maketitle

\begin{abstract} 		
If spikes are the medium, what is the message? Answering that question is driving the development of large-scale, single neuron resolution recordings from behaving animals, on the scale of thousands of neurons. But these data are inherently high-dimensional, with as many dimensions as neurons - so how do we make sense of them? For many the answer is to reduce the number of dimensions. Here I argue we can distinguish weak and strong principles of neural dimension reduction. The weak principle is that dimension reduction is a convenient tool for making sense of complex neural data. The strong principle is that dimension reduction shows us how neural circuits actually operate and compute. Elucidating these principles is crucial, for which we subscribe to provides radically different interpretations of the same neural activity data. I show how we could make either the weak or strong principles appear to be true based on innocuous looking decisions about how we use dimension reduction on our data. To counteract these confounds, I outline the experimental evidence for the strong principle that do not come from dimension reduction; but also show there are a number of neural phenomena that the strong principle fails to address. To reconcile these conflicting data, I suggest that the brain has both principles at play.


\end{abstract}

\section{Introduction}
Neurons communicate moment-to-moment using spikes. Many believe that capturing as many spikes from as many neurons as possible is a promising route to understanding the brain. In principle, such data will contain all the messages we need to know about. And large-scale, single neuron resolution recordings are now available from many neural circuits in a range of species, from hundreds of neurons in a literally detached retina watching a film \citep{Tkacik2014}, to thousands of neurons in the visual cortices of mice \citep{Stringer2019,Stringer2019a}, to tens of thousands across the brain of a baby zebrafish \citep{Ahrens2012,Portugues2014,Vladimirov2018}.

But these long-sought data are revealing a new challenge. The joint activity of a large neural population is both complex and high dimensional: it has as many dimensions as neurons, and each neuron's activity traces a unique pattern over time. So if we are to use such data to advance our understanding of the brain, first we have to solve the problem of understanding the data.

For many, the solution is to turn to dimension reduction \citep{Cunningham2014,Pang2016}. These analytical tools find the components of activity that co-vary across the members of a neural population. Roughly speaking, when applied to data on neural activity, dimension reduction aims to replace the many individual sequences of activity from each neuron with a few sequences of activity that each describe the common patterns found across many neurons.

I propose here that we should distinguish weak and strong principles of neural dimension reduction. The weak principle is that dimension reduction is a convenient tool for making sense of complex neural data. The strong principle is that dimension reduction shows us the true latent signal(s) encoded by a population of neurons, and so moves us closer to how neural circuits actually operate and compute. Which principle we subscribe to provides radically different interpretations of the same dimension reduction techniques applied to the same data.

\subsection{Neural dimension reduction}
\begin{figure*}  
	\centering
	\includegraphics{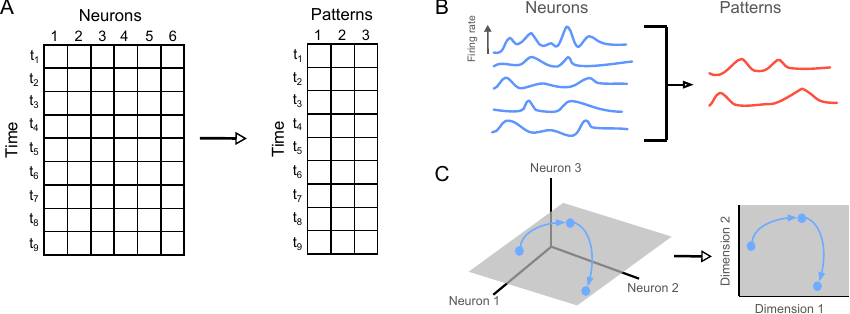}
	\caption{\label{fig:dimreduction} {\bf Sketches of neural dimension reduction. } \\
		A: Dimension reduction in matrix form. The activity of $N$ neurons is captured in a matrix $\mathbf{A}$ with as many rows as time-steps, each entry the activity of neuron $i$ at time-step $t$. Dimension reduction applied to the columns -- the time-series of neural activity -- aims to reduce the dimensions of the matrix, to find the $d \ll N$ most common patterns of activity of time. \\
		B: Dimension reduction of neural activity, a schematic. Given the output of five neurons over time (blue), applying dimension reduction reduces these to the most common sequences of activity shared by the neurons -- here, two (red). \\ 
		C: Dimension reduction as finding a low-dimensional space of joint activity. We can think of the joint activity of three neurons as living in a three dimensional space, one dimension per neuron (left). Plotting their firing rates as they evolve over time (arrows) creates a trajectory of activity. The activity of the neurons co-varies such that it always remains in the grey plane. Which means that we need just two dimensions to capture the variation in activity of these three neurons (right): so dimension reduction applied to the time-series also finds us this low-dimensional space (with caveats -- see main text).
	}	
\end{figure*}

We are considering here dimension reduction applied to the time-series of many simultaneously recorded neurons. Given $N$ neurons recorded for $T$ time-steps, we create an $T \times N$ matrix $\mathbf{A}$ that encapsulates the recorded population -- one column per neuron, one row per time-step. Neural dimension reduction thus aims to collapse $\mathbf{A}$ to a new matrix $\mathbf{P}$ that is $T \times d$, where the number of new dimensions $d$ is ideally much less than the number of neurons $N$. Each column of $\mathbf{P}$ is interpreted as a sequence of activity that is common across many neurons (Figure \ref{fig:dimreduction}). 

\edit{A key question for dimension reduction is}: how big should $d$ be? If the dynamics of a neural population are simple, in that most neurons contribute to a handful of common sequences of activity, then $d$ can be small compared to $N$; if the dynamics are complex, with few common sequences between neurons, then $d$ will be large compared to $N$. \edit{How we determine and then interpret $d$ forms the basis for all that follows} \footnote{An alternative form of neural dimension reduction is to find the common patterns of activity across all the neurons in a population. That is, we keep all neurons $N$, and instead reduce the number of time-points $T$: we take matrix $\mathbf{A}$, and reduce it to a matrix $\mathbf{S}$ that is $d_t \times N$. As each row of $\mathbf{A}$ is the pattern of joint activity over the population at that time-point, so reducing it to $\mathbf{S}$ is finding the $d_t$ most common of those population activity patterns \citep{Gao2017}. We expect $d_t$ to scale too: if a population has simple dynamics, often revisiting similar patterns of joint activity, then $d_t$ will be small compared to $T$; with complex dynamics, meaning few common joint activity patterns, then $d_t$ will be large relative to $T$. While the discussion here is framed in terms of reducing the number of dimensions compared to the number of neurons, many of the following arguments for weak and strong principles apply equally well to reducing the number of population states too.}.

\section{The weak and strong principles} 
Let's begin with the obvious question: Why make this distinction between weak and strong principles of neural dimension reduction?

The weak principle is that dimension reduction is an interpretative tool \citep{Cunningham2014,Pang2016,Jazayeri2017}. By taking us from $N$ neurons to $d$ dimensions, it provides us with a way of collapsing these $N$ sequences of activity, up to tens of thousands, into $d$ sequences of activity, typically by factors of 10 or more \citep{Gao2015}. Each of the $d$ sequences is a composite of the individual sequences, capturing the things they have in common. And it removes ``noise'', because by describing the $N$ neurons in terms of the activity common to two or more neurons, so the fluctuations unique to each neuron are eliminated. Such ``noise'' is variability that is not controlled by the experimenter, whether intrinsic ``noise'' like variation in the response of a single neuron over repeated exposures to the same stimulus \footnote{Intrinsic ``noise'' is noise from the observer's point of view, not the brain's. \edit{When we eliminate (many) higher dimensions as noise, we inevitably run the risk of removing elements of neural activity that could be crucial for understanding the coding or computation of the neural population. For example, population activity projected into a low dimensional space is unlikely to contain a meaningful contribution from any ``soloist'' neurons \citep{Okun2015} as by definition their activity is independent from the majority. Recent modelling work \citep{Sweeney2020} suggests that in visual cortex such soloist neurons are those with the least variable stimulus responses, and thus by eliminating them dimension reduction would potentially eliminate the most consistent response to a given stimulus}.}, or measurement noise from, say, noise in sensor fluorescence in calcium imaging. The simplified and cleaner representation allows us insight into the coding or computation by the whole population we have recorded.

Under the weak principle, we can use dimension reduction to find the information available across a group of neurons. We look at the $d$ sequences of activity we end up with, and ask what they encode; or what we can decode from them. Or ask which neurons contribute the most to each of the $d$ dimensions, so working out which neurons respond similarly. 

The weak principle is then that dimension reduction is a useful tool, sometimes fantastically so, but nothing more, because it does not reveal to us anything fundamental about how the brain works. It is useful because we only get to observe the brain's activity over a short window of time relative to the brain's whole lifespan, so we can describe the here-and-now in relatively few dimensions. And this is very useful if we want to try and get our heads around the brain; but is not the claim that the brain really does operate with far fewer dimensions than neurons.

The strong principle is that claim: dimension reduction shows us the true underlying signal embodied by the neural circuit. The so-called ``latent signal''. It is a theory that the brain really is low-dimensional, compared to the number of neurons. That the joint activity of a population of neurons is a (noisy) realisation of this low-dimensional system -- a realisation using many more elements (neurons) than dimensions in the system.

Under the strong principle, it is the low-dimensional trajectory of activity that encodes information \citep{Briggman2005,Shenoy2013,Kato2015,Bruno2017,Gallego2017}. One trajectory for swim; another for crawl. One for go left; another for go right. One for reach up, one for down. At its most extreme, the strong principle says that any coding we see in single neurons is an epiphenomenon of the joint coding by the population. So when we find single neurons that fire just before a mouse turns left, it is not because the neuron itself is ``tuned'' to moving left, but because it contributes most to the trajectory that means ``left''.

Put another way, the strong principle is that what we're seeing in the brain is a $d$-dimensional dynamical system implemented by $N$ individual elements, where $N$ is much greater than $d$ \edit{\citep[for more on a dynamical systems view see e.g.][]{Vyas2020}}. Why not just use $d$ neurons? Because neurons are fragile and synapses are unreliable, so degeneracy is needed -- the loss or failure of one neuron or of one spike cannot crucially disrupt things. And as neurons transmit using spikes, so each can only approximate the continuous dynamics encoding a dynamical system. Hence, ``noisy'': a population of $N$ neurons implements a $d$-dimensional dynamical system by simultaneously solving the constraints of transmitting reliable signals and robustness to damage.

\section{What is the dimensionality of neural activity?}
Naively, differentiating the weak and strong principles should be easy: apply your dimension reduction technique of choice to your activity matrix $\mathbf{A}$ and see how many dimensions $d$ we need to retain to capture most of the variation of activity. (In classic principle components analysis (PCA), we do this by simply checking how much additional covariance of the data is accounted for by each added dimension). Few dimensions relative to the size of the population is consistent with the strong principle; many dimensions is consistent with the weak principle.  

Many studies have used dimension reduction techniques on population recordings, but curiously few systematically explore the dimensionality of their data. Prior studies that have compared the number of dimensions in neural activity have mostly focussed on trial-averaged responses \citep{Churchland2007a,Machens2010,Lehky2014,Cowley2016,Gao2017}. For example, \citet{Lehky2014} estimated the dimensions of single neuron coding in inferior temporal (IT) cortex, by first taking the mean spike count of each of 674 neurons in response to the presentation of each one of 806 stimuli (their matrix $\mathbf{A}$ was thus 674 neurons by 806 mean responses). Applying dimension reduction to trial-averaged data is thus asking about the representation space -- how many dimensions span the space of single neuron tuning (to stimulus, to memory, to movement). In their IT cortex data \citet{Lehky2014}, for example, report they need 7.9\% of all possible dimensions to account for the shared responses of 674 neurons to the 806 stimuli; extrapolating to a population with infinite neurons they estimated a total capacity of about 100 dimensions. Such attacks on representation space are deeply interesting questions, but not quite what we're after here: the dimensions of ongoing activity, the dimensionality the brain gets to work with in the moment.

Ongoing activity in invertebrate systems seems low-dimensional. During the \emph{Aplysia}'s escape gallop, we found just 5 to 8 linear dimensions ($\sim$ 5\% the size of the recorded population) is needed to account for 80\% of the variance between neurons in its motor system; and adding more dimensions did not improve the decoding of motor output \citep{Bruno2017}. Similarly, \citet{Briggman2005} report low numbers of linear dimensions are needed to separate the trajectories of ongoing activity that correspond to swimming and crawling in the leech's motor system. Yet even here, this work only indirectly tackles the question of dimensions of neural activity, over a tiny snapshot of time (a few seconds), in a single behaviour.  

These examples show that simply asking for the dimensionality $d$ of ongoing activity is fraught with potential misunderstandings. We'd need to define our task limits; the brain state of interest; and where we put the dividing line between low and high dimensions.  

To some, defining neural dimensionality requires long recordings of a neural population exposed to a rich set of stimuli (to probe everything in the world they care about) or during a rich set of movements (to probe everything in the body they care about). Simple tasks or stimuli may only exercise a neural population over but a few of the dimensions it can reach, masking a high-dimensional system \citep{Gao2015,Gao2017}. So before we take a measurement of $d$, and argue whether it supports the weak or strong principle, we need to define our task limits: dimensions for one movement, or task, or set of stimuli; or all of them? 

And in what brain states? After all, the dimensionality of a brain region in resting, behaving, REM and non-REM sleep are all likely different. Apparent low dimensions in the spontaneous activity of a population of V1 neurons in the anaesthetised macaque \citep{Williamson2016a} is likely simply because most anaesthetics produce highly correlated activity in the form of up/down transitions \citep{Ecker2014}, that would then be read-out as low-dimensional shared activity across a population. So measuring $d$ in order to support the weak or strong principles also needs us to define the brain states we are interested in. 

Simply measuring dimensionality $d$ is also not enough, for what defines ``high'' or ``low'' is in the eye of the beholder. An example: recently, \citet{Stringer2019a} simultaneously imaged around 10,000 neurons in mouse V1 during spontaneous movement for an hour or more. The corresponding spontaneous neural activity during this hour was highly structured, with reliable correlations between neurons, and with the dominant components of population activity being self-correlated over time-scales of tens of seconds. Using a new approach to look at the dimensions of activity reliably shared across the population over time, \citet{Stringer2019a} reported that 128 of these ``shared variability'' components accounted for about 86\% of the population's variance. While this was interpreted in the paper as being evidence for a high-dimensional latent signal, I note that this amounts to roughly 1\% of all possible dimensions for this population, and a factor of hundred drop in dimensions compared to neurons could be interpreted as low-dimensional. 

But let's say we agree our terms for a given brain area: the brain state, the length of recording and complexity of task, and even what divides $d$ into ``high'' or ``low'' given those terms. Even then, establishing $d$ may be analytically challenging: there are many ways we can confound the weak and strong principles just by how we handle the data.

\section{Confounds of dimensionality}
Using dimension reduction to directly establish whether the strong principle ($d \ll N$) or weak principle holds for a population would require showing that the result is not because one accidentally chose an inappropriate way of representing the data or of processing the data. It is also challenging because we have to make many assumptions about the state of the neural population being analysed. Let's examine some of these confounds.

\subsection{Nonlinearity}
Classic dimension reduction techniques, such as PCA, are linear. If we apply a linear method to neural activity data and keep $d$ dimensions, then we are assuming the neural activity sits on a flat $d$-dimensional plane. But the actual shape -- the manifold -- on which the neural activity sits could be a curved surface, could be nonlinear. 

\begin{figure*}  
	\centering
	\includegraphics{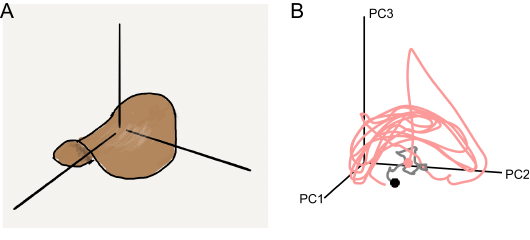}
	\caption{\label{fig:pringle} {\bf Nonlinear dimensions} \\
		A: Schematic example of intrinsic and embedding dimensions -- the curved manifold has two intrinsic dimensions on its surface, but occupies an embedding volume of three linear dimensions. \\
		B: Real example of a manifold that is apparently not linear in three dimensions. The trajectory shows the joint activity of 105 neurons in the \emph{Aplysia} motor system from spontaneous (grey) to evoked fictive movement (pink). The bulk of the evoked trajectory describes a curved two-dimensional surface in the three-dimensional visualisation. 
	}
\end{figure*}

Which means we need to separate two different types of dimensions, the embedding and intrinsic dimensions \citep{Camastra2003}. The intrinsic dimensions are the number of dimensions needed to describe the surface; the embedding dimensions are those needed to describe the space occupied by the surface. If the surface is a plane, then the intrinsic and embedding dimensions agree - they are both two (see Figure \ref{fig:dimreduction}C). But say a population's neural activity sits on a surface shaped like a popular curved potato-based snack (Figure \ref{fig:pringle}). Then it has two intrinsic dimensions - a ``pringle'' is a two dimensional shape - but three embedding dimensions, because the ``pringle'' occupies a three-dimensional volume. Thus the embedding dimensions are the upper limit of the true intrinsic dimensions of the population activity.   

Linear methods for dimension reduction can only recover these embedding dimensions. The $d$ dimensions we have kept are the upper limit of dimensionality: they are the embedding dimensions needed to fully capture the shape of the surface on which the activity actually sits. So the true dimensions -- the intrinsic dimensions -- of the activity could be much lower. This means that if we find a high dimensional space using linear dimension reduction, it is not evidence against the ``strong'' principle; but finding a low dimensional space with linear methods is evidence for it.

An example: above I suggested that \citeauthor{Stringer2019a}'s \citep{Stringer2019a} recording of spontaneous activity from about 10,000 neurons in the mouse V1 could be interpreted as low-dimensional: their estimate of 128 dimensions is about 1\% of the maximum possible. That estimate was reached using a linear method for dimension reduction, and so is actually the upper limit of the dimensionality of the activity. It is the embedding dimensions: the intrinsic dimensions of the population activity could be considerably lower.  

\edit{There are some neural systems where we reasonably expect the intrinsic dimensions of their activity to be well-defined and irreducibly low. One clear example is the head-direction system, the network of neurons whose activity keeps track of the current heading angle of the animal, with reference to some landmark. We have excellent evidence that heading direction in \emph{Drosophila} is encoded by neurons in their ellipsoid body that form a ring attractor \citep{Kim2017}. These neurons sustain a persistent bump of activity within a ring of neurons that represent the current heading direction \citep{Seelig2015}. The hypothesised existence of a ring attractor strongly implies that the intrinsic dimensions of the neural activity are one-dimensional, moving only around a loop that continuously encodes the 360 degrees of possible heading directions. More recent work on the thalamic regions of the head direction system in mice has indeed provided compelling evidence that the joint population activity within the anterodorsal thalamus falls on a loop that has one intrinsic dimension, but potentially many embedding dimensions \citep{Chaudhuri2019}.} 

\edit{It is unclear whether there exist other neural systems whose intrinsic dimensions are so well-defined. Systems that generate cyclical movements are one set of candidates \citep{Bruno2017}. Nonetheless, techniques to find the intrinsic dimensions, and in some cases directly model the manifold of activity within them, have been applied to neural activity data, including the correlation dimension \cite{Grassberger1983,Lehky2014}, Isomap \citep{Tenenbaum2000,Mimica2018}, Laplacian eigenmaps \cite{Belkin2003}, UMAP \cite{McInnes2018,Tombaz2020} and persistent homology \cite{Singh2008,Chaudhuri2019}. For example, when \citet{Singh2008} used persistent homology to study the spontaneous and evoked activity in groups of 5 neurons in macaque V1, they found tantalising hints that the group's activity fell on a sphere. Robustly demonstrating the existence of a manifold of population activity with few intrinsic dimensions would be considerable evidence for the strong principle.}

\subsection{The tough problem: time}
Hidden in the above is a small but not innocuous assumption. We started with the idea that we want to apply dimension reduction to the sequences of activity of $N$ neurons. Our starting point is the matrix $\mathbf{A}$, $N$ columns wide and $T$ rows long, each column describing the activity of one neuron in our population. But that means we have to divide up that activity into $T$ discrete blocks, each block a time-step of size $\delta T$. What should the time-step be? \footnote{Also crucial is what we represent in each discrete time-step. In calcium imaging, each time-step is (at least) one frame, but its contents could be the fluorescence time-series, or inferred rates of spiking, or even inferred spikes. For electrophysiology, the entries will be binary (spike or not), spike counts, or some transform thereof.}.

\begin{figure*}  
	\centering
	\includegraphics{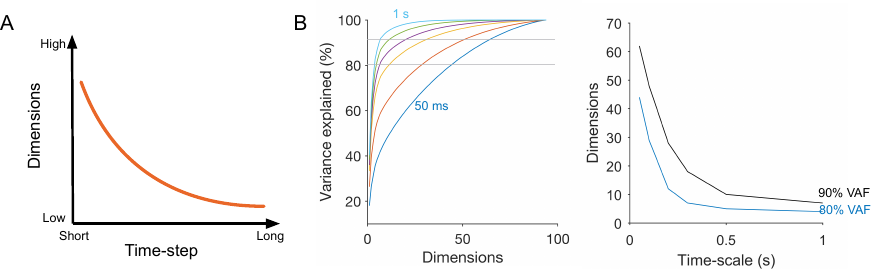}
	\caption{\label{fig:timeconfounds} {\bf Effects of time-step on neural dimensionality} \\
		A: Schematic of the dependence of dimensionality $d$ of neural population activity on the underlying time-step of the neural activity matrix. We expect the dimensionality to fall as we increase the length of the time-step from milliseconds to seconds, as we are changing the requirement for correlations between the neurons in the population. \\
		B: Dimensionality of population activity in the \emph{Aplysia} motor system falls as a function of time-step. I applied PCA to a population of 95 neurons recorded from the  \emph{Aplysia} pedal ganglion at the onset of a fictive escape gallop. Left: the cumulative variance explained by each additional dimension, for each time-scale used to define the population activity. Right: the scaling of dimensions with time-scale, for dimensions needed to account for 80\% or 90\% of the total variance in the population (grey lines in the left panel).  
	}	
\end{figure*}

Generally, the smaller we make $\delta T$, the more precise we are asking correlations between neurons to be. (Indeed, large $\delta T$ implies a rate code and small $\delta T$ implies a spike-timing code). And precise correlations are rare. The smaller we make $\delta T$, the lower the apparent correlations between neurons, and fewer the common patterns of activity between neurons. Consequently, the smaller we make our time-step, the higher number of dimension $d$ we get from our dimension reduction (Figure \ref{fig:timeconfounds}).  

Without doing anything else, we can alter $d$ by orders of magnitude just by changing the time-step $\delta T$. So we can seemingly support either the weak or strong principles by choosing the time-step size to fit our prejudices \footnote{\edit{Smoothing our time-series of calcium fluorescence or spikes will also change the apparent correlations between neurons -- the wider the time-window used to smooth, the likely higher the correlations will be: Changing the representation of activity will also change the apparent dimensionality of that activity}}.

Defining the time-step $\delta T$ is then a crucial decision in establishing the dimensionality of your data. Its size is often fixed by some aspect of the experimental set up. Imaging experiments have a fixed frame-rate, as low as 2 frames per second, so it is natural to just use one frame per time-step. And that is often the implicit decision. But there is no necessary relation between a meaningful time-step for the neural dynamics of the circuit being recorded and what the frame-rate happens to be. (Indeed, a persistent worry with imaging experiments is that the frame rate is too slow to capture some key aspects of neural activity). 

To illustrate, let's return to the example of \citeauthor{Stringer2019a}'s \citep{Stringer2019a} recording of spontaneous activity from about 10,000 neurons in the mouse V1. Above I suggested these neurons' activity could be interpreted as low-dimensional, because the number of dimensions needed to recover a large proportion of the original activity was a small fraction of the number of neurons. But these spontaneous activity data were calcium imaging time-series sampled at a 2Hz frame-rate. So these dimensions were defined on a time-step of 500 ms, very long on the time-scale of individual spikes, which I've just argued will inevitably give us a low estimate of $d$. One could then argue that $d$ on the scale of spikes in the spontaneous activity of V1 may indeed be high-dimensional. 

The \citet{Stringer2019a} data have thus shown us the interplay of these confounds of dimensionality. With these same data I have been able to argue that their 128 dimension estimate is a lower bound, because the time resolution is so low, and at the same time an upper bound because the dimensions are linear and so only describe the embedding space. Playing with how we represent and reduce our data lets us find the answer we want. 

\edit{To circumvent this confound of time}, we'd like to define the time-step by the characteristic timescale of whatever is reading out the population. Rarely do we have some idea of what this is; or even if it's a meaningful question to ask. For motorneurons projecting to muscles, perhaps, when we have some idea of the time-scale of the parameters of the movements. Indeed, whenever we have behaviour available, we can use that to at least provide upper limits on what the time-step should be -- for it has to be shorter than the changes in behaviour. Even with this information, the time-step needn't even be fixed, of course. The time-scale at which neural activity is read-out is likely flexible.

\edit{As we noted in the last section, we expect there to exist neural systems with a low intrinsic dimension: if so, we also might expect them to be more resistant to this confound of time, because that structure should exist at all time-scales. But all methods for finding the dimensions of experimental data are based on some measure of distance between data-points, be it correlation, Euclidean distance, or something else \citep{Belkin2003,McInnes2018}. Changing $\delta T$ changes these distances between pairs of neurons (or between vectors of the entire population's activity). Ideally, a low-dimensional structure would be found across many choices of $\delta T$ if all the distance relationships between points on this structure are preserved and simply scaled up or down; but we likely add more noise as we progressively make $\delta T$ smaller, and so alter the ordering of distances.}

Being unable to objectively fix the time-step of neural activity means we do not know \emph{a priori} at what time-scale exists the ``true'' dimensionality of the activity of a neural system, or even if one can be defined. We must make some choice, or explore a range of choices. An interesting program of work would be to look at how this scaling of dimensions with time-scale depends on the dimension reduction methods we use and on the species and brain region examined. Hence we need to be aware of the weak and strong principles, to know that choosing the time-scale we use to describe our data must colour our interpretation of them \footnote{\edit{Indeed, one might argue that a key unsolved problem in systems neuroscience is a reference frame of time: what time-scale(s) of neural output are relevant to behaviour, and what time-scale(s) are relevant to the brain's internal dynamics}}.

\subsection{Confounds of the neural activity itself}
The above are all confounds in the process of analysing the neural data. But say we could solve all of them, then we still have confounds of dimensionality that arise from the inherent properties of neural activity.

Applying any dimension reduction approach assumes the dimensionality of the population activity is stationary \citep{Perich2018a}. Indeed, any studies that use trial-averaged responses as a basis for dimension reduction assume at the outset that the dimensions of the population activity are stationary. There are good reasons to suspect they are not. One is that neuromodulation of a neural population will change the effective connections between its neurons on short time-scales, and so change the resulting population activity \citep{Katz1996,Marder2012}. Another is that learning will also change the connections between neurons, on equivalent or slightly longer time-scales. Anything that alters the effective connections between the neurons in a population could alter the dimensions of its activity. Whether such changes are sufficient to switch from an evidently low to high-dimensional population activity, or vice-versa, is an open question.

\section{Support for the strong principle beyond dimension reduction}
\edit{That we can find a low-dimensional representation of neural activity does not of course mean that the brain uses it, as claimed by the strong principle. An adherent of the weak principle could still posit that the low-dimensional representation is a mere epiphenomena of an alternative theory of neural representation.} And all the above confounds may make us question any dimensionality estimate. But further support for the strong principle comes from the convergence of independent, individually suggestive lines of evidence. Here are some of those.

\paragraph{Neurons are correlated.}
The sneaking suspicion that population activity is low dimensional arises from us long knowing that the activity of neurons is correlated in time \citep{Wohrer2013}. It is correlated during quiet waking \citep{Ringach2009}, during spontaneous behaviour \citep[e.g.][]{Stringer2019a}, and across the presentations of different stimuli \citep{Averbeck2006} (so-called ``noise'' correlations; signal correlation being the correlation of a pair of neurons' average response to a stimulus -- see \citealt{Cohen2011} for definitions and discussion). Even weak correlations between its neurons implies that a population visits just a small subset of the possible states -- patterns of joint activity -- that would be visited by independent neurons \citep{Schneidman2006,Ganmor2015,Singh2019a}. The mere existence of correlation implies population activity must have fewer dimensions than neurons, because some of its activity is redundant: there are points in time where pairs (or more) of neurons in that population have approximately the same activity.

Finding that such correlations are stable over long periods of time is consistent with the existence of some kind of low-dimensional system realised by that population. Such long-lasting correlations have been reported for populations of grid cells \citep{Yoon2013}, and of head-direction cells across waking and sleep \citep{Peyrache2015}, populations of the \emph{Aplysia} motor system across an hour or more \citep{Bruno2017}, for noise correlations in layer 2/3 of primary visual cortex \citep{Jeon2018}, and for spontaneous activity in primary auditory cortex over days \citep{Betzel2019,Ponce-Alvarez2020}. 

\paragraph{The same population of neurons drive two or more qualitatively different behaviours.} The invertebrate literature calls these ``multifunctional" neurons, neurons whose activity correlates with (or, better, causes) two qualitatively different behaviours \citep{Briggman2008}. A canonical example is the ganglion neurons of the leech that participate in both swimming and crawling. If we extend the definition of ``behaviour" to include the global dynamics of the circuit, then this also includes, for example, neurons of the crustacean stomatogastric ganglion, which supports two globally different rhythms (pyloric, gastric), but some neurons are common to both \citep{Marder2007}. 

A simple explanation for such multifunctional circuits is that the population of neurons implement (at least) two different low-dimensional attractors (each attractor can have arbitrarily complex dynamics), and something switches the circuit from one to the other \citep{Briggman2008}. In invertebrates, we know that something is likely to be a neuromodulator \citep{Katz1996,Marder2012}.

\paragraph{Neural ensembles are well isolated.} A vast literature is devoted to the idea of the neural ensemble, a group of neurons that are consistently co-active and so likely computing or coding the same thing \citep{Harris2005,Yuste2015,Holtmaat2016}. Typically such ensembles are found by clustering time-series (i.e. grouping the columns of $\mathbf{A}$). While simple clustering (with e.g. k-means) will always return groups, one can use more sensitive approaches, with null models, that will detect well-isolated groups of co-active neurons \citep{Humphries2011,Lopes-dos-Santos2011}. By definition, finding $E$ ensembles where  $E \ll N$ is also support for the strong principle: for it means there is considerable redundancy between neurons \citep{Bruno2015}.

\paragraph{Knocking out neurons does not kill the dynamics and/or function of a circuit.} If a neural circuit's dynamics are high dimensional, then knocking out a few neurons should have a measurable effect: after all, some dimensions have been lost. But in a number of neural circuits, we see that destroying a few neurons has little to no effect on its dynamics (and, by extension, function, whatever that may be). Indeed, optogenetic experiments routinely have to use sledge-hammer levels of stimulation to kill a circuit's dynamics \citep{Otchy2015} -- whether via directly inhibiting spiking or by exciting GABAergic interneurons \citep{Wiegert2017}. Even the neural activity and behaviour of the 302-neuron nematode worm \emph{C Elegans} is robust to having a couple of its neurons incinerated \citep{Kato2015}.

\paragraph{Downstream decoding is of the latent signal.} A further clue to support the strong principle for a neural population would be if one could show that its downstream targets make use of its low-dimensional latent signal. We showed a little of this in our study of the dynamics of the crawling circuit in the sea-slug \emph{Aplysia}: that circuit's low-dimensional latent signal is sufficient to decode the commands being sent to the neck muscles, and the decoding is not improved by adding more dimensions of activity \citep{Bruno2017}. Similarly, \citet{Pandarinath2018} showed they could decode the kinematics of a center-out arm reach from a learnt low-dimensional representation of population activity in primary motor cortex, with fantastic performance (their Figure 4). 

\paragraph{Constraints on neural plasticity are low dimensional.} If a neural population's dynamics were high dimensional, then this implies they could also change along many dimensions. But if low dimensional, then the changes would likely be constrained to those dimensions. Sadtler and colleagues reported some evidence for this in the monkey's motor cortex \citep{Sadtler2014}. They tasked a monkey with controlling a cursor using just the activity of a neuron population of the motor cortex. The challenge lay in how they mapped from the population activity to the cursor movement. Each day, they first mapped the low-dimensional space occupied by the population's ongoing activity \footnote{\edit{They showed the joint activity of approximately 90 recorded units took an average of 10 linear dimensions to fully capture}}, and found the directions along which the activity mapped to the movement of the cursor. Then they changed this mapping between activity and the cursor: in one condition, let's call it ``aligned'', they changed the mapping within the low-dimensional space; in the other condition, they mapped the cursor movement to axes rotated outside the low-dimensional space. \edit{The ``aligned'' condition was much easier: keeping the problem within the low-dimensional space made re-learning the mapping faster, and achieved more accurate control, consistent with a low-dimensional encoding of movement}. 

\edit{In further work, \citet{Golub2018} showed that the patterns of joint activity that made up the low-dimensional space did not change before and after learning in the ``aligned'' condition. Consequently, it seemed the monkeys learnt the remapping between the neural activity and the direction of movement not by aligning the population's activity to the new mapping, but by changing what the existing activity patterns meant -- which in turn implies that the relearning was done by changing the inputs to the recorded population that coded for the intended direction of movement. Thus not only were the changes within the low-dimensional space easier to learn, there was no apparent change of the low-dimensional activity at all.} The next, more compelling, step would be to show that ``natural'' plasticity is also constrained to low-dimensional activity.

\paragraph{Synaptic turnover does not alter a circuit's dynamics.} There is a growing body of evidence that properties of synapses, like spine size, change spontaneously \citep{Ziv2017}. Turnover in these properties changes the effective strength of connections between neurons, so changing the excitability of neurons and changing who excites them. An open question then is: how does a circuit keep a stable output to support its functions in the face of this constant change? (An assumption here is that a circuit's function needs a stable output of any kind). The strong principle is one answer: small shifts in the excitability of individual neurons would not affect the low-dimensional latent signal of a population. \edit{And indeed we now have evidence, for example, that the low-dimensional signals in primate motor cortex that correspond to different directions of reaching can remain stable across time-scales of weeks to years \citep{Gallego2020}}. 

\section{Challenges for the strong principle}
The above is a demonstration of consilience, the convergence of multiple individually suggestive lines of evidence around a single hypothesis. But there are key neural phenomena that are challenging to explain under the strong principle. Here are some of those. 

\paragraph{Sparse coding}
The theory of efficient or sparse coding predicts that in sensory regions of cortex the neural responses to stimuli are sparse across the population and in time \citep{Wohrer2013}. It predicts a sparse population response because only a few neurons will respond to a given stimulus, those most precisely tuned to its features, and this precise tuning also means that each neuron will only respond sparsely over its lifetime, activated only by the rare occurrences of the specific feature(s) it is tuned to. Sparse coding thus implies a very high-dimensional code when considering an entire cortical sensory region, such as V1: the specificity of each neuron's tuning means there is little shared variance between the activity of neurons. 

In a direct test of sparse coding on this scale, \citet{Stringer2019} characterised the dimensionality of the space containing the responses of a population of about 10,000 neurons in mouse V1 to 2800 natural images. They showed the response-space is indeed high dimensional, such that the 2800 population responses do not sit in any clearly defined subset of the 2800 possible dimensions \footnote{The above confounds are still potentially in play here: these are linear dimensions, so an upper limit of the embedding space of the population activity; and assessing the dimensionality over 2800 images assumes stationarity over about an hour of recording.}. Though these were crude measurements of the static responses to each stimulus (two frames of a deconvolved calcium signal per neuron), their results suggest that the pattern of activity across a population in mouse V1 in response to visual stimuli is not low dimensional, and thus inconsistent with the strong principle. 

Nor do their results support the ideas of sparse coding either. Pure sparse coding theories predict that each dimension of population activity is approximately of equal importance, because no combination of neurons should be more consistently co-active than any other. But \citet{Stringer2019} report the dimension's importance in V1 scaled as a power law, meaning some combinations of neurons were more consistently co-active than others. If anything, their results support a model of V1 activity as medium-dimensional. 

And this nicely illustrates the broader issue that sparse coding ideas pose for the strong principle: on what spatial scale is the population low-dimensional? If we record nearby V1 neurons during a stimulus, we likely capture some responding to the same stimulus and so see co-active neurons \citep{Cossell2015}, a potentially low-dimensional population. But if we record a large fraction of V1, then most neurons are not active together, and so the population activity would likely be of far higher dimensions. Thus the larger the region sampled, the likely the greater heterogeneity of coding in our population, and so the higher the apparent dimensionality of its activity.

\paragraph{Cell types} 
Implicit in the strong principle is the idea that if the population's dynamics are carrying information, then cell types are not important. This is most explicit in recurrent neural network approaches to analysing population dynamics \citep{Sussillo2014}, where one either replicates or fits the low-dimensional dynamics of a neural population with a recurrent network model \citep[e.g.][]{Mante2013,Hennequin2014,Pandarinath2018}. These recurrent networks have no cell classes, beyond the existence of inhibitory and excitatory neurons; indeed some don't even follow Dale's law of having solely one signed type of neurotransmitter per neuron, freely mixing inhibitory and excitatory output from the same neuron. Thus, we can easily replicate the low-dimensional dynamics of the cortex without reference to different classes of cell within it. 

Yet clearly cell types are rife. The latest detailed survey of mouse cortex using single-cell transcription sequences gave 117 different classes of neuron, 56 expressing glutamate, the vast majority being types of pyramidal neuron, and 61 expressing GABA, likely all interneurons \citep{Tasic2018}. We see the same diversity in human cortex, with all 69 neuron types detected by transcription sequencing in a sample of human neurons matching known types in mice \citep{Hodge2019}. Classifications that include electrophysiological responses and connection targets on top of genetically-defined types could be broader still. And it has long been obvious, from Cajal onwards, that different brain regions have their own unique sets of neurons, often seemingly exquisitely designed for the task at hand \citep{Stiefel2007,Cuntz2010}. 

Such diversity is unnecessary according to the strong principle: if regions of the brain are computing using low-dimensional dynamics, and we can get these dynamics from any basic recurrent network, we don't need cell type diversity. Why then, for example, exists the chandelier cell in mammalian cortex, a GABAergic interneuron that targets the axon initial segment of a pyramidal cell, specifically so that it can suppress the release of a spike without affecting dendritic activity. An answer might be that the chandelier cells are crucial to the homeostatic regulation of activity across a local cortical circuit \citep{Pan-Vazquez2020}. The challenge for adherents to the strong principle is to explain why cell type diversity exists: whether they are epiphenomena, or essential to endowing a population with the necessary low-dimensional activity. 

\paragraph{Dendritic computation}
Similar issues arise when we look closer at the individual neuron. It has long been established that pyramidal cell dendrites can support complex computations, including logic operations, sequencing, and coincidence detection \citep{Koch1983,London2005,Branco2010,Jadi2012}. One mechanism is through local ``spikes'' in the dendrites, which allow non-linear summation of inputs in apical dendrites \citep{Polsky2004}. Such spikes allow a single pyramidal cell to function as a two-layer neural network \citep{Poirazi2003a}. And even the passive properties of dendrites allow computation of an extended range of non-linear functions within a single neuron \citep{Caze2013}.

Such dendritic computation is \edit{a puzzle for adherents to the strong principle, for they have no need of that hypothesis. As I noted above, from recurrent networks we can seemingly obtain any form of low-dimensional dynamics we desire (indeed, recurrently connected networks of simple point neurons are a universal approximator of any dynamical system that can be written in a discrete time, state-space form \citep{Schaefer2007}). And these recurrent networks use a cartoon model of a neuron, one that linearly sums its inputs and passes them through a nonlinear, typically sigmoidal, output function. Thus if the brain encodes information in the low-dimensional activity of a neural population, then neurons in those populations likely do not need computations within their own dendrites in order to create those low-dimensional dynamics. Yet apparently they do have the capacity for such computation. Again, a challenge for adherents of the strong principle is to explain why dendritic computation exists, and what it is for.}

\paragraph{Precise spike timing in the periphery}
The strong principle is a population doctrine, not a neuron one, but it makes some implicit assumptions about the apparent ``code'' used by the spikes of individual neurons. Spike-timing codes predict that repeating the same stimulus would evoke the same pattern of spikes from a single neuron, with minimal jitter. A simulated recurrent network would of course repeat the same low-dimensional trajectory given noiseless dynamics, identical starting conditions, and an identical input. But reality is likely different: any low-dimensional dynamics created by cortical-like recurrent circuits are unlikely to repeat so precisely that individual neurons within the circuit repeat the same spike patterns \citep{Banerjee2008}, because identical inputs and starting conditions do not happen.  

Yet mammalian neurons are capable of such spike-time precision, especially neurons receiving direct input from sensory receptors, such as the ganglion cells of the retina \citep{Berry1997}. At the most extreme, \citet{Bale2015} report neurons in the trigeminal ganglion can respond to the repeated movement trajectory of a whisker with a spike-time precision on the order of tens of microseconds ($\mu$s). Spike-time delays on the same order are crucial to the decoding of the angle to a sound source in the owl's hearing circuit \citep{Carr1990}. And while the precisely timed spikes in these neurons are many synapses removed from cortex (at least three in the case of the ascending whisker pathway), one wonders why such precision is necessary if the end result is for them to be washed away in local recurrent activity of cortex \citep{Ringach2009}. Similarly confusing is that millisecond changes in the spike times of spinal motorneuron firing can alter movement, at least in insects and songbirds \citep{Sober2018}. Another  puzzle then: why input precisely timed spikes to a system using low-dimensional dynamics? \footnote{One suspects the answer is precisely because these are inputs to recurrent activity in cortex: that precisely timed inputs across neurons in the early sensory system provides the synchronised input to cortex necessary to perturb the ongoing dynamics \citep{Bruno2006}}  And how and why output them?

\section{What is not evidence either way}
There are many other aspects of neural activity that at first glance seem to speak to either the weak or strong principle, but on deeper reflection do not. I briefly summarise these here, and give a fuller account for interested readers in the Appendix. No doubt some of the above list will join these in time. 

Single neurons with mixed tunings, or populations with no apparent single neuron tuning could be interpreted as a consequence of a low-dimensional latent signal. But this is affirming the consequent: the existence of a low-dimensional latent signal means neurons could seem to encode conjoint features of the world, or nothing at all, depending on how their activity contributed to that signal. But the inverse is not true: mixed encodings can be high-dimensional \citep{Rigotti2013} (they can, after all, be sparse too), and the absence of tuning might simply mean we've not been smart enough to work out what those neurons do encode. 

Other phenomena that do not speak either way to the strong or weak principle include reports that single spikes affect behaviour, that individual neuron's firing can vary dramatically between repeats of the same movement, and that neurons in cortex can reproduce precise spike times (where precise is on the order of tens of milliseconds, so orders of magnitude more than at the periphery). Again, all these can be explained under either principle: that a single spike is effective may seem to say individual neurons are important and so inherently high-dimensional, but a single spike can alter the trajectory of an entire population; variably-active neurons across repeats of the same movement could be either correlated (low dimensional) or uncorrelated (high dimensional); and spike timing in cortex is not so precise as to rule out population-wide fluctuations in firing rate.  

\section{Lessons}
What lessons might we draw from the idea of strong and weak principles of neural dimension reduction?

\subsection{Directly comparing the weak and strong principles}
One lesson is the need to develop approaches that directly pit the two principles against each other. I sketch three ideas here using dimension reduction, with the acknowledgement that all remain susceptible to the confounds of time-scale, nonlinearity, and stationarity discussed above. \edit{And, of course, with the caveat that simply showing a neural population is low-dimensional is necessary, but not sufficient, for the strong principle.}

\begin{figure}  
	\centering
	\includegraphics{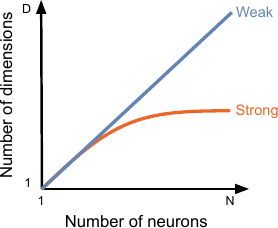}
	\caption{\label{fig:scaling} {\bf Differentiating the weak and strong principles using scaling} \\
		Schematic of the expected scaling in the number of dimensions with the number of neurons included in a population that would differentiate the weak and strong principles. The number of dimensions is that needed to capture some fraction $F$ of the covariation in the $N$ neurons' activity.
	}	
\end{figure}

The first idea is that we measure how the dimensions of a population's activity depend on its size. Say we start with a population of $N$ neurons, whose activity needs $d$ dimensions to capture a fraction $F$ of their total activity. As we add more neurons from the same population, so the number of dimensions $d$ needed to capture $F$ percent of variance must stay the same or grow (Figure \ref{fig:scaling}). That is, if we grow the population by $(N+1,N+2,\ldots,N+m)$ and keep $F$ fixed, then $d(N+m)$ must be bounded between $d$ and $d+m$. The rate of growth of $d$ gives us a quantitative deciding criterion between the two principles for that specific population: if $d$ plateaus and does not grow with more neurons, then this is consistent with the strong principle; if $d$ continues to grow, this is evidence against the strong principle, and so consistent with the weak principle. 


The second idea is to use the data itself to find the number of dimensions. This could be done using cross-validation, by splitting the data into train and test sets. We then fit a dimension reduction model to the training set, reconstruct the testing set using the model by choosing some number of dimensions $d$, and find $d$ which optimises the reconstruction (for examples of cross-validated PCA, see \citep{Wold1978,Eastment1982,Diana2002}). The resulting $d$ gives us an idea of the true ``latent'' dimensions. \edit{That said, there are many difficulties. Cross-validating such population time-series data would need sufficiently long time-series for stable estimates of the held-out portion of the data, which would likely grow with the number of neurons $N$. In practice, such cross-validation can have no clear peaks, but rather show a plateau of possible dimensions (see e.g. Figure 4a of \citep{Sadtler2014}). And cross-validation of course assumes that the data are stationary across the testing and training sets: but with such long time-series of neural activity this is a strong assumption.}

The third idea is to use a null model for the expected dimensionality of the data, in order to define ``low dimensional''. \edit{There are at least two ways of doing this. We can create null models by synthesising time-series that contain key features of the original time-series data but that control or eliminate the correlations between those time-series that give rise to low-dimensional structure. Independently shuffling the individual time-series to eliminate correlations between them is simplest; more advanced models that preserve data features are available \citep{Elsayed2017}. Alternately,} many dimension reduction methods work by starting with a comparison matrix $\mathbf{C}$, of pairwise comparisons between rows or columns of $\mathbf{A}$. PCA, for example, uses the covariance matrix as $\mathbf{C}$. For these methods, the dimension question can be reframed as how many dimensions we'd need to reconstruct $\mathbf{C}$. So one could define a null model $\mathbf{C}_{\text{null}}$ for what $\mathbf{C}$ would look like if it had no low-dimensional structure, and determine the number of dimensions $d$ of $\mathbf{C}$ that depart from the null model \citep{Humphries2021}. 

\subsection{The past, present, and future of systems neuroscience}
Historically, systems neuroscience's approach to neural coding was to record individual neurons, find ones tuned to a stimulus or movement, and concentrate all analysis and subsequent theories on those \citep{Wohrer2013,Hardcastle2017}. The strong principle implies this historical approach was wrong. If the low-dimensional signal of a neural population is what a brain works with, then any single neuron tuning is uninformative about how that brain computes. Tuning arises because that neuron consistently participates in the part of the low-dimensional latent signal that encodes property X of the world, be it orientation, direction, frequency, amplitude, pressure, rotation or other elementary features we correlate with neuron activity. And more complex properties too: place cells in the hippocampus can be explained as just the neurons that participate in their population's trajectory at a particular point in space \citep{Rubin2019}. If the strong principle is true our historical fixation on tuning is misleading at best. 

The perspective of the weak and strong principles also lets us reinterpret swathes of recent systems neuroscience studies according to their implicit theories of how the brain works. Any study which records many neurons at the same time during some task or stimulus set is implicitly subscribing to the weak or strong principle. Studies that analyse the activity of each individual neuron are implicitly subscribing to the weak principle. Studies that analyse a projection of the individual neuron activity implicitly subscribe to the strong principle. And that implicit principle will colour all other aspects of the study, especially how the results are interpreted.

And what of the future? If the strong principle is true, then the goal for systems neuroscience should be to find the underlying dynamical system for a given population. So we can find what a population of neurons is encoding, even if -- especially if -- encoding is not readily apparent in single neurons \citep{Maggi2018,Zylberberg2018,Insanally2019,Stringer2019}. So we can understand how that population solves the computational problems it faces, such as how similar activity driving the same behaviour in the present can evolve into distinct activity driving different behaviours in the future \citep{Russo2018,Russo2020}.  So we can compare activity between the same circuit in different conditions \citep{Gallego2018}; so we can compare activity in the same circuit between different animals \citep{Bruno2017}.

We can view the strong principle as the answer to the following conundrum: how does a population of neurons perform the same functions across time, tasks, or even brains? Over time, wiring and neurons change. And any two brains from the same species, even the tiny brains of \emph{Drosophila} and the simple molluscs (leeches, \emph{Aplysia}, \emph{Lymnaea}), have different wirings, different ratios of types of neurons, and different single neuron dynamics. The strong principle says that the same function arises because the latent dynamics of the population remain the same: the low-dimensional space is robust to variations in individual neuron properties, across tasks \citep{Gallego2018}, time \citep{Hill2015,Bruno2017,Gallego2020}, and brains \citep{Bruno2017}. 

This conservation of function offers a compelling question for future research programmes of how a population's low-dimensional latent signals map onto the information encoded in the population, on to the latent \emph{variables}: how many latent variables are encoded by a population, whether that number is fixed or learnt, and whether the number and type of latent variables are consistent across brains of the same species.

Above I listed issues for the strong principle. One view is that these issues are evidence for the weak principle. Another view is that they constitute a research programme: if the brain does encode latent signals in the low-dimensional activity of a neural population, then why do cell classes, dendritic computation, and spike-time precision exist? An even larger research programme looms: from the strong principle it follows that the low-dimensional latent signals are what learning acts on, are what evolution sculpts.

Indeed, an ongoing programme of research by Srdjan Ostojic and colleagues is constructing a theory for why low-dimensional dynamics exist in the brain, a theory that suggests the strong principle may be true because neural circuits find it easier to operate in low dimensions. They've shown that low-dimensional dynamics in recurrent neural networks can be guaranteed by the existence of a low rank (i.e. low dimensional) structure to the weights between neurons, embedded in otherwise unstructured connections \citep{Mastrogiuseppe2018}. Remarkably, finding appropriate low-rank embedded weights allows a recurrent neural network to carry out a wide range of cognitive tasks -- including parametric working memory and context-dependent two-alternative decision making tasks -- and do so with only one or two dimensions of activity, created by just one or two effective populations of neuron within the network \citep{Dubreuil2020}. Moreover, the act of searching for network weights that will implement a cognitive task causes low-rank structure to appear within the weights \citep{Schuessler2020}. Thus, the apparent low-dimensional activity of a neural population might simply be because the brain's network of synaptic connections inevitably ends up with an embedded low-dimensional structure itself when changing with experience.

\edit{\subsection{Two thought experiments}}
\edit{The idea of the strong and weak principles suggests two interesting thought experiments. I outline these and the insights they give, and leave the reader to draw their own conclusions}.

\edit{The first is what we would learn if we repeatedly assessed the dimensionality of an identical population of neurons under identical recording and analysis conditions, what we might call the ``relative'' dimensionality. Much of the above discussion has been about how confounds alter our estimates of \emph{absolute} dimensionality, a single number that is arrived at by a given study -- examples are given above of dimensions of activity in mouse V1, primate IT cortex, and the \emph{Aplysia} motor system. Our list of confounds suggest such absolute estimates are fragile, only interpretable within the context of the system they are established in and the conditions under which they are determined. By holding the confounding factors, such as time-scale, fixed, measuring relative dimensionality ought to bring us more insight into whether a neural population follows the strong principle. A starting hypothesis would be that, if the dimensions of neural activity are largely determined by the wiring of the underlying neural circuit, then we should expect to find those dimensions are of a consistent magnitude every time we measure them. In particular, if the strong principle is true, then if we find $d \ll N$ once for a population then we might expect to find it always under the same analysis assumptions. Though, again, of course that will be defined by the time-scales at which we analyse those time-series of neural activity.}

\edit{The second thought experiment is if we extrapolate to being able to record every neuron in a brain for as long as we wanted: Would this provide the definitive answer to whether the brain operates as a low-dimensional system or not? Finding many fewer dimensions than neurons, and finding this is consistent across time, is suggestive but lacks causal evidence that the brain uses this low-dimensional representation. However much we record, we need causal manipulation. One such manipulation is to stimulate or ablate neurons, and demonstrate that changes to the low-dimensional activity have a causal effect, whether that be within the brain or on behaviour. \citet{Briggman2005} give a beautiful example of this approach: in a segmental ganglion of the leech, they found individual neurons whose activity strongly differed depending on whether stimulating the ganglion evoked fictive swimming or crawling; yet hyperpolarising or depolarising those strongly-tuned neurons had no effect on whether the evoked behaviour was swimming or crawling. By contrast, when they identified a different neuron that contributed most to separating the low-dimensional activity underlying the swimming or crawling behaviour, they found that manipulating it strongly biased the consequent behaviour towards swimming, when it was hyperpolarised, or crawling, when it was depolarised. Evidence, then, that altering the low-dimensional trajectory of activity caused a change in the evoked motor program. Yet even with the current plethora of causal tools at our disposal, interpreting their results remains tough. Stimulating or ablating groups of neurons can have ``off-target'' effects, where those neurons are not themselves causal for the behaviour, but manipulating them alters dynamics elsewhere in regions of the brain that are \citep{Otchy2015,Hong2018,Wolff2018}. So simply finding that manipulating low-dimensional activity alters behaviour need not, unfortunately, tell us anything about the strong principle.}

\edit{The idea of extrapolating to the scale of the whole mammalian brain brings into sharp focus a further issue: that the number of neurons we record can be dwarfed by the number of actual neurons in any given brain region. While imaging 10,000 neurons simultaneously in mouse V1 \citep{Stringer2019,Stringer2019a} is breathtaking, there are about half a million neurons in its V1 \citep{Herculano-Houzel2013}. And while we have gained fantastic insights by reducing recordings of tens of neurons in macaque motor cortex to 6 \citep{Churchland2012} or 10 \citep{Sadtler2014,Russo2018} dimensions, there are about 50 million neurons in its primary motor cortex (M1) alone \citep{Turner2016}. Even if we find the strong principle holds for such relatively small populations, would it still hold when scaled to entire brain regions? Even if it does, it might not help: a reduction in the number of dimensions by a factor of hundred compared to the number of neurons is still 5,000 dimensions in mouse V1, and 500,000 dimensions in macaque M1. What we hope is that my sketch above in Figure \ref{fig:scaling} is true: that the number of dimensions we find in our small populations does not scale linearly, so that we are already actually close to the number of dimensions in the entire population of interest.}



\subsection{The brain has both}
We as scientists might implicitly subscribe to one or the other. Might the brain subscribe to both the weak and strong principles?

\begin{figure*}[!]  
	\centering
	\includegraphics[width=0.5\textwidth]{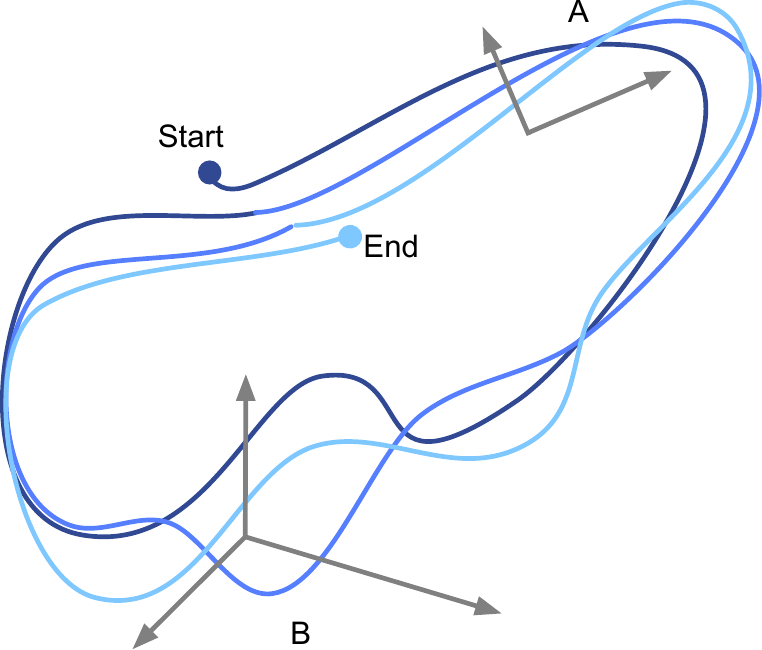}
	\caption{\label{fig:both} {\bf Both weak and strong principles can be simultaneously true in the brain} \\
		I sketch here the trajectory of activity of every neuron in some small, hypothetical ``whole'' brain, of unknown, arbitrary number of neurons $N$; the trajectory shows three loops over time, from dark to light blue. Imagine we record in two places in the brain, marked A and B. \edit{If we record in A, then we can project that set of trajectories onto a two-dimensional plane: the trajectories all fall in two dimensions, so touch when they cross. But if we record in B, then the projection will have (at least) three dimensions, as the trajectories separate along all three axes and so do not touch at any point}. 
	}	
\end{figure*}

Some neural systems seem to have strong redundancies in their activity, and work as low-dimensional systems. There is evidence for low-dimensional constraints on activity in a range of invertebrate systems -- particularly in the motor systems of the leech \citep{Briggman2005} and \emph{Aplysia} \citep{Bruno2015,Bruno2017}, and in the whole nervous system of \emph{C Elegans} \citep{Kato2015,Nichols2017,Brennan2019}. There is evidence from different regions of cortex, including primary motor cortex \citep{Gallego2018}, posterior parietal cortex \citep{Raposo2014}, and auditory cortex \citep{Bartho2009}. And there is evidence from the head direction system in flies \citep{Seelig2015,Kim2017} and mammals \citep{Zhang1996,Peyrache2015,Chaudhuri2019}. Unsurprisingly, the strongest evidence for true low-dimensional systems comes from systems that control low-dimensional movements or encode low-dimensional properties of the world.

But other neural systems must represent a multitude of different things, an unknown number, and so need a rich capacity for coding. And these constraints favour, but do not guarantee, a weak principle. The visual system in mammals is the canonical example. The statistics of the visual world are complex, and never static. From this flow compelling arguments for high-dimensional activity -- for efficient codes that minimise redundancies in neural activity as far as possible in the face of competing constraints of error correction and energy use \citep{Stringer2019}.

Visual, parietal and motor cortices are all part of the same mammalian brain, implying that the brain subscribes to both the weak and strong principles of neural dimension reduction. The unanswered question is then: how can one transform into the other? Answers to this are obviously dimension reduction from weak to strong; and dimension expansion from strong to weak. There are some nice theories of how this would work \citep{Ganguli2012}, but all are framed around compression from $N$ neurons to a smaller group of $M$ neurons; or of expansion from $M$ neurons up to $N$ neurons. But the brain is not a collection of discrete layers, from which one can expand or compress into another.

For I hid an assumption in plain sight above: that we want to know if the dynamics of a ``circuit`` or a ``network'' adhere to the weak or strong principle. But what brain circuit is isolated from all others? None. A brain is one massive recurrent dynamical system. So we cannot just lift out one small area of cortex and claim it high dimensional and another small area and claim it low dimensional, and expect that to be the answer, because ultimately they are just part of one massive sheet of interconnected neurons, connected to and from a massive ball of neurons underneath them.

So what does it mean if one small area of cortex appears to subscribe to the weak principle and another to the strong? One answer is that they are transformed \citep{Perich2018a,Semedo2019}. But another is that our microscopic snapshots of the giant $n$ dimensional attractor that is the brain are sometimes of low dimensional parts of that attractor and sometimes of high dimensional parts (Figure \ref{fig:both}). If the brain subscribes to both the strong and weak principle of neural dimension reduction, so ought we.

\section*{Acknowledgements}
This idea grew from writing the Discussion section of \citep{Bruno2017} on attractor dynamics of the \emph{Aplysia} motor system -- I thank Angela Bruno and Bill Frost for our time on that paper, and the sample data used here. My lab (Silvia Maggi, Mat Evans and Francois Cinotti) and Nikos Gekas gave great feedback on these ideas at lab meetings. Discussions with attendees of the NORDITA Workshop on Dimensionality Reduction and Population Dynamics in Neural Data (Stockholm, February 2020) firmed up and altered some of these ideas, and added a few new ones. The need to separate embedding and intrinsic dimensions has been repeatedly stressed in talks by Srdjan Ostojic. Finally, I thank Juan Gallego and Juan Galeazzi Gonzalez for their comments on drafts of this manuscript, and the three anonymous reviewers for their constructive comments that refined the paper -- particularly for prompting the discussion of systems where we expect a low-dimensional manifold to exist, and of the two thought experiments. 

\section*{Conflict of interest}
The author declares no conflicts of interest.


\bibliography{WeakStrongRefs}

\begin{thebibliography}{127}
\bibitem[{Hardcastle et~al.(2017)Hardcastle, Kiah and Ganguli, Surya and
  Giocomo, Lisa M}]{Hardcastle2017}
Hardcastle K, Ganguli S, Giocomo LM.
\newblock Cell types for our sense of location: where we are and where we are
  going.
\newblock Nature Neurosci 2017 Oct;20:1474--1482.
\end{thebibliography}

\appendix
\section{Appendix}
\subsection{What is not evidence for either the weak or strong principles} 
\paragraph{Multiplexed single neuron tuning.} Single neurons in a range of cortical areas have mixed tunings -- they respond to multiple thing happening in the world with no obvious ``types'' \citep{Wohrer2013,Rigotti2013,Raposo2014,Maggi2018}. Such conjoint coding could suggest a low-dimensional code. But a population could equally have a high-dimensional representation of mixed tunings, where neurons are uniquely tuned to their particular combination, and so act independently. Indeed \citet{Rigotti2013} explicitly argue that one reason for mixed tuning -- what they call ``mixed selectivity'' -- is to deliberately create a high dimensional representation from a low dimensional one, because doing so can map a nonlinearly separable problem to a linearly separable one. (That is, they envisage projecting a low-dimensional nonlinear separation problem into a higher dimensional space to make it linearly separable by a hyperplane). But this requires nonlinear mixed tuning: the response to a combination of things is not some linear sum of individual responses to those things.

\paragraph{Accurate decoding from a neural population, but not single neurons of that population.} We and others have shown that some cortical regions have coding of stimuli or events that is barely perceptible at a single neuron level, but is clear when decoded from the larger combined population \citep[e.g.][]{Maggi2018}. Again this might be taken as evidence for the strong principle: that we need the conjoint activity of many neurons. But no. It could equally imply a distributed code, where individual neuron firing varies across the population according to stimulus (or event), but the neurons need not consistently co-vary. Whether such population-only decoding supports the strong or weak principle would depend on the details of how that coding manifests.

\paragraph{Single spikes affecting behaviour.} If the addition (or deletion) of spikes in a single neuron can detectably alter some aspect of an animal's behaviour, then this might seem strong evidence for the weak principle -- that the circuit at hand does not read out from the joint activity of many neurons. For example, \citet{Houweling2008} reported that adding spikes to a single neuron in the somatosensory cortex is sufficient for a rat to detect a change in its neural activity and start licking. However, we are well versed by now in the tricky nature of causality in neuroscience. Adding or deleting a few spikes from a single neuron can be enough to change the trajectory of an entire cortical population \citep{Izhikevich2008,Li2009a,London2010}.

\paragraph{Individual neurons vary trial-to-trial, but the population activity does not change.} Neurons can be fickle things. Sometimes a neuron responds strongly to a stimulus, or fires strongly during a movement. Sometimes the same neuron can't be coaxed into firing at all. Bill Frost and colleagues have shown in both \emph{Aplysia} and \emph{Tritonia} that an entire motor circuit can be made to cleanly repeat a set of rhythmic dynamics, wave of bursting activity across neurons to drive rhythmic behaviour, and yet individual neurons in that circuit vary dramatically in how much they participate -- both between repeated bursts within the same motor program \citep{Hill2012}, and between repeats of the entire motor program \citep{Hill2015,Bruno2017}. That a circuit doesn't need the same neurons to do the same thing would seem evidence that the circuit is encoding its key information in some low dimensional form. But there could equally be unreliable neurons in a high-dimensional code: if their unreliability is not correlated then the same neurons are rarely active together, meaning the population as a whole has high dimensional dynamics. Thus support either way would depend on exactly what type of unreliability is in play.

\paragraph{Spike-time precision on repeated trials in the cortex.} If we repeat a stimulus we can ask if the neuron(s) repeat the same spikes at the same moment in time. Observing precisely repeated spikes in individual neurons would seem at odds with the encoding of a low-dimensional signal in the population, because precise spikes require the whole population's trajectory to be repeated with high accuracy. 

We have reports of spike-time precision in cortex: in area MT in response to clouds of randomly moving dots \citep{Bair1996} and time-varying stimuli \citep{Buracas1998}, and areas of IT in response to images \citep{Amarasingham2006}, among others. But ``precise'' in these studies means a jitter between trials of at least 10 ms per spike, orders of magnitude greater than at the periphery. Moreover, the precision fades quickly, with spikes beyond a few hundred milliseconds after the stimulus onset less aligned across trials. These details are consistent with the visual stimulus repeatedly triggering the same feedforward input signal to these neurons, causing the evolution of dynamics from a similar starting point, and with that evolution following a Poisson process with a rapidly time-varying rate to obtain a jitter on the order of 10s of milliseconds per spike.

Moreover, there are good theoretical arguments for why a recurrently connected network like the cortex cannot ever use spike time precision, as it would require the network to converge on the same low-dimensional trajectory from a wide range of initial conditions, and for that trajectory to be repeated precisely \citep{Banerjee2008,London2010}. That said, this still leaves open the question: why then do so many peripheral systems seem to use precise spike timing?

\end{document}